\documentclass[%
 reprint,
 amsmath,amssymb,
 aps,
prb,
]{revtex4-2}

\usepackage{graphicx}
\usepackage{dcolumn}
\usepackage{bm}

\begin{document}

\preprint{v:1.0~~~~~15.8.20}

\title{A search for non-reciprocal magnons in MnPS$_3$}

\author{A. R. Wildes}
 \email{wildes@ill.fr}
\affiliation{Institut Laue-Langevin, 71 avenue des Martyrs CS 20156, 38042 Grenoble Cedex 9, France}

\author{S. Okamoto}
\affiliation{Materials Science and Technology Division, Oak Ridge National Laboratory, Oak Ridge, Tennessee 37831, USA}

\author{D. Xiao}
\affiliation{Department of Physics, Carnegie Mellon University, Pittsburgh, Pennsylvania 15213, USA}

\date{\today}

\begin{abstract}
Recent articles have suggested that the quasi-two dimensional antiferromagnet MnPS$_3$ may have non-reciprocal magnons, whereby  magnons in a Brillouin zone corner at $+${\bf{q}} have different energies than those at $-${\bf{q}}.  The magnons along the Brillouin zone boundaries were measured using neutron three-axis spectrometry, paying careful attention to the resolution function, to determine whether such non-reciprocity was present.  The data show that, within the resolution, there are no significant differences between the magnons in opposite Brillouin zone corners.
\end{abstract}

\maketitle


MnPS$_3$ belongs to a family of layered van der Waals compounds that have attracted considerable attention \cite{Brec, Grasso,Susner,Wang}.  The van der Waals nature of the compounds gives them physicochemical properties that have been studied as possible candidates for optical sensors and battery materials, and even cancer treatments \cite{Mohamad}.  More recently, the ability to delaminate the compounds has been explored which has attracted the attention of the graphene community \cite{Park}, especially as a number of members of the family are intrinsically magnetic.    

MnPS$_3$ is one of the magnetic family members.  It has a $C~2/m$ space group, with the $S = 5/2$ Mn$^{2+}$ ions forming a honeycomb lattice in the \emph{ab} planes \cite{Ouvrard85}.  The compound orders antiferromagnetically below its N{\'e}el temperature of $\sim 78$ K, \cite{Okuda86, Joy92} adopting a {\bf{k}} = 0 collinear structure where each ion is antiferromagnetically coupled to its three nearest-neighbours \cite{Kurosawa83}.  The moments are almost normal to the \emph{ab} planes, tilted by $\sim 8^{\circ}$ towards the $a$ axis \cite{Ressouche}.  Its paramagnetic susceptibility is isotropic, showing that the compound has a Heisenberg-like Hamiltonian.

The possibilities to use magnetic layered compounds in graphene technology requires the understanding of their spin dynamics.  A number of theoretical studies have considered the spin dynamics in a collinear antiferromagnet on a honeycomb lattice.  The Mermin-Wagner theorem states that an isotropic Heisenberg Hamiltonian will not give rise long-ranged magnetic order in two dimensions \cite{Mermin}, and extra terms need to be added to the Hamiltonian to stabilise any ordered magnetic structure.  Added terms have included dipole-dipole anisotropy \cite{Pich95}, a Dzyaloshinskii-Moriya interaction \cite{Cheng}, and a bond-specific anisotropic exchange \cite{Matsumoto}.  

\begin{figure}
   \includegraphics[width=3in]{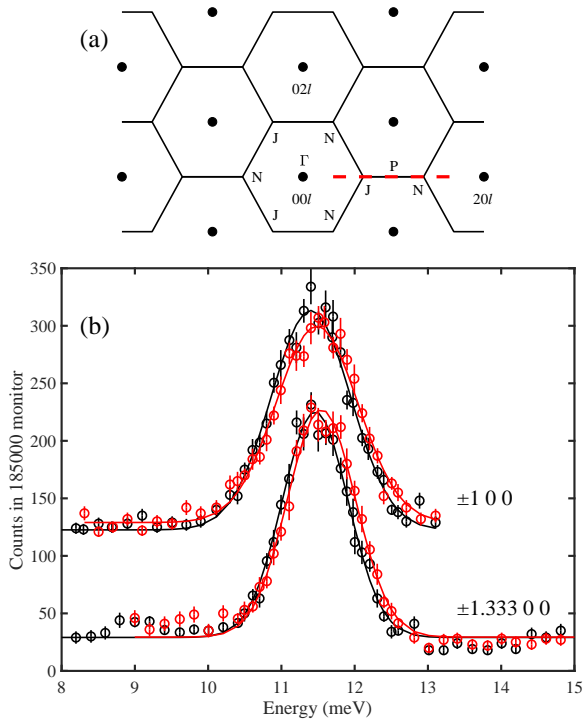}
   \caption{\label{fig:Data} (a) Schematic showing a plane in reciprocal space for the magnetic structure of MnPS$_3$.  A Brillouin zone centre is marked as $\Gamma$, and high-symmetry points on the Brillouin zone boundary are marked with $J$, $N$ and $P$.  The sample was mounted in the $h0l$ plane for the experiments, which is orthogonal to the plane shown here.  The experiments focused on the scattering along the $h00$ and $h01$ directions, shown by the dashed red line. (b). Data measured at $\pm\frac{4}{3}00$, corresponding to $J$ and $N$ corners of the Brillouin zone, and $\pm100$, corresponding to $P$ points.  The measurement time for each data point was $\sim 330$ seconds.  The +{\bf{Q}} and $-${\bf{Q}} data points are shown in black and red respectively.  The data at $\pm100$ have been shifted vertically by 100 for clarity.  Fits of Gaussians, including a flat background, to the data are also shown.}
\end{figure}

These three theories predict non-reciprocal magnons, where magnons at reduced scattering vectors of $\pm${\bf{q}} have different energies.  The differences are greatest at the Brillouin zone boundaries.  Figure \ref{fig:Data}(a) shows a reciprocal lattice plane for the honeycomb lattice with the Brillouin zone boundaries indicated.  The theories for the dipole-dipole anisotropy and bond-specific exchange predict that the non-reciprocity takes the form of a splitting in the two-fold degeneracy of the antiferromagnetic magnons, and that the splitting is different at reduced scattering vectors of $\pm${\bf{q}}.  The splitting is greatest at the corners labelled $J$ and is zero at those labelled $N$.  The mean energies for the magnons, however, is the same at both points.  The theory for the Dzyaloshinskii-Moriya interaction, however, predicts that the magnons stay degenerate, but the dispersion becomes asymmmetric about the $\Gamma$ point with the magnons at $J$ having different energies to those at $N$.  The last theory has particular interest for graphene technology as it may lead to a spin Nernst effect, where magnons with a selected chirality could be excited and driven using a temperature gradient \cite{Cheng}.  It is worth searching for physical representations of such a system to test the theory, and  MnPS$_3$ was specifically presented as a candidate that may have non-reciprocal magnons \cite{Pich95, Cheng, Matsumoto}.

Neutron three-axis spectrometry has been previously used to study the spin dynamics of bulk MnPS$_3$ \cite{Wildes98}.  Although the {\bf{k}} = 0 antiferromagnetic structure gives rise to two magnon modes, only one could be detected within the instrumental resolution.  The dispersion surface had an energy gap of 0.5 meV at the Brillouin zone centre and rose to $\sim 11.5$ meV at the Brillouin zone boundary.  The surface was fitted using a Heisenberg Hamiltonian with an easy-axis term for the anisotropy.  Satisfactory fits required the inclusion of three in-plane nearest-neighbours. The interplanar exchange was $\approx 1/400$ the magnitude of the first nearest-neighbour exchange, showing that bulk MnPS$_3$ is a good approximation of a two-dimensional magnet. 

The anisotropy was small, being $\approx 1.1\%$ the magnitude of the first nearest-neighbour exchange, and its origin is somewhat ambiguous.  At least two effects are believed to contribute: 

A significant contribution appears to arise from dipole-dipole anisotropy, which would lead to a splitting in the degeneracy of the two magnon modes that is greatest at the $J$ points and is zero at the $N$ points \cite{Pich95}.  The theory for a bond-specific anisotropic exchange leads to the same effect \cite{Matsumoto}.  Such a splitting was confirmed using neutron spin-echo spectroscopy \cite{Hicks}.  The splitting at $J$ was 64 $\mu$eV, which is very small and well within the resolution of the three-axis spectrometer measurements.  The measured splitting and the spin wave gap were significantly smaller than expected from calculations if dipole-dipole interactions were the sole source of the anisotropy, which would furthermore lead to the moments pointing normal to the $ab$ planes \cite{Goos10} rather than being tilted towards the $a$ axis \cite{Ressouche}.

A single-ion anisotropy is also believed to be present, based on electron spin-resonance experiments on dilute Mn in CdPS$_3$ \cite{Okuda86}.  The anisotropy was determined to lie in the $ab$ planes, which is consistent with the tilting of the moments and would help to explain the observation that the critical properties of the magnetism in MnPS$_3$ map onto an XY-like Hamiltonian \cite{Wildes06}.  The presence of a single-ion anisotropy is somewhat unexpected given that a free Mn$^{2+}$ ion has no orbital angular momentum and its source is not known.

The neutron data to date would suggest that, aside from the very small splitting most likely due to dipole-dipole anisotropy, the magnons in MnPS$_3$ are symmetric either side of the Brillouin zone centre.  However, the measurements did not explicitly test for this.  This article therefore presents a dedicated search for non-reciprocal magnons using neutron three-axis spectrometry, paying special attention to the instrumental resolution.   

Neutron three-axis spectrometry on MnPS$_3$ was carried out using the IN8 spectrometer at the Institut Laue-Langevin, France \cite{Wildes_IN8_Jun19}.  The instrument was configured with a pyrolytic graphite (PG) $002$ monochromator and analyser, which were horizontally flat and vertically focused on the sample.  The horizontal divergences were limited using $40^{\prime}$ collimators before and after both the monochromator and the analyser.  The final wavenumber was fixed at $k_f = 2.662${\AA}$^{-1}$, and higher order wavelength contamination was suppressed using a PG filter between sample and analyser.  The same MnPS$_3$ crystal used in previous neutron studies \cite{Ressouche, Wildes98, Wildes06, Hicks} was aligned such that $\left(h0l\right)$ was the scattering plane, and the sample was cooled to 1.8 K using a liquid helium cryostat.

The experiments consisted of energy scans at constant {\bf{Q}}, focusing on the magnetic scattering along the $h00$ and $h01$ directions.  As indicated by the red dashed line in figure \ref{fig:Data}(a), this trajectory includes a Brillouin zone boundary with access to the $J$, $P$ and $N$ points.  The figure also shows that a $J$ point at scattering vector $+{\bf{Q}}$ is matched by an $N$ point at $-{\bf{Q}}$.  The instrumental resolution is a function of the magnitude of the scattering vector, $Q$, and the energy transfer, $\hbar\omega$.  Corresponding $J$ and $N$ points were measured by rotating the sample by $180^{\circ}$ about the normal to the scattering plane, thus allowing a direct relative comparison and minimising potential systematic errors associated with a different resolution.  Similar measurements were performed at the $P$ points.

Figure \ref{fig:Data}(b) shows data measured at $\pm\frac{4}{3}00$, corresponding to $J$ and $N$ points, and $\pm100$, corresponding to $P$ points.  The data show clear peaks from magnons that are approximately dispersionless with energy $\hbar\omega \approx 11.5$ meV, consistent with the previous measurements that showed spin waves with the same energy and with very little dispersion between $0.5 \le h \le 1.5$ \cite{Wildes98}.  The peaks disappeared in measurements above the N{\'e}el temperature, verifying their magnetic origin.  The data were fitted with Gaussians to give the peak centres and full-width half-maxima (FWHM), and the fits are shown in the figure.  Similar measurements were performed at corresponding points for $l = 1$, and the fit results are summarised in table \ref{tab:Results}.

\begin{table}
   \begin{tabular}{cc|cc|cc}
   			&	& \multicolumn{2}{c|}{Centres ({\AA}$^{-1}$)} & \multicolumn{2}{c}{FWHM ({\AA}$^{-1}$)} \\
	$\left|h\right|$ & $\left|l\right|$ &$-${\bf{Q}}	&	{\bf{Q}}	    & $-${\bf{Q}}	 &	{\bf{Q}} \\
	1.333             & 0		     &  11.55(1)	&      11.46(1)	    & 1.09(3)	 &   1.06(3) \\
	1.333	     & 1		     &  11.62(2)   &	11.43(2)	    & 1.05(6)	 &   1.09(5) \\
	1		     & 0		     &  11.49(1)   &	11.40(1)	    & 1.32(4)	 &   1.28(3) \\
	1		     & 1		     &  11.43(4)	&     11.32(3)	    &  1.26(9)	 & 1.24(7) \\	
   \end{tabular}
   \caption{\label{tab:Results} Results from fitting Gaussians to the neutron scattering data at different points on the Brillouin zone boundary.  The calculated energy resolution for a dispersionless mode at 11.5 mV is 0.9145 meV.}
\end{table}

An initial inspection of the peak centres at $\left|h\right| = \frac{4}{3}$ shows a systematic difference of $0.09$ meV between $\pm${\bf{Q}}.  However, this must be tempered by the observation of a similar difference between the centres at $\left|h\right| = 1$.  The theories all show that the mean energies for the magnons at the $P$ positions should be equivalent \cite{Pich95, Cheng, Matsumoto}.  Thus the small energy differences, which are $< 9${\%} of the FWHM, are likely to be an experimental artefact, possibly due to the centre of mass of the sample being slightly off the central axis of rotation for the spectrometer.  It must therefore be concluded that the measurements show no significant difference between the energies of the magnons at the $J$ and $N$ points. 
 
The possibility that the data at $\pm\frac{4}{3}00$ in figure \ref{fig:Data}(b) contain, in fact, multiple peaks that are not resolvable within the instrument resolution must be considered.  The presence of crystallographic and antiferromagnetic domains may cause the scattering at these positions to consist of the magnons at both $J$ and $N$ points, and thus the experimental data would be a superposition of three peaks for the dipole-dipole and bond-specific theories, and two peaks for the Dzyaloshinskii-Moriya theory.  The crystal structure of MnPS$_3$ is monoclinic, but it is very close to being orthohexagonal \cite{Ouvrard85} and, being layered, it is prone to stacking faults and twinning.  The twins have a distinct relationship, corresponding to a rotation by $120^{\circ}$ about the {\bf{c}}$^{\star}$ axis \cite{Murayama}.  This rotation in itself would not map J points onto N points, however such a mapping may occur when combined with stacking faults \cite{Ouvrard90}.   The antiferromagnetic domains are equivalent to a rotation of the honeycomb lattice by $180^{\circ}$ about the normal to the plane which explicitly maps the J points onto the N points.  The antiferromagnetic domain population will depend on the way that the sample is cooled.  It is possible to drive MnPS$_3$ into a monodomain state by cooling the sample in crossed electric and magnetic fields \cite{Ressouche}, but this was not done during the experiment.  It is noteworthy that the same sample was shown to be essentially single-domain in the neutron spin-echo experiment [17].  The possible influence of domains can be tested comparing the fitted FWHM in table \ref{tab:Results} with the expected instrumental resolution.



Consequently, careful attention was paid to the resolution function for the instrument.  An experimental estimate for the {\bf{Q}}-resolution was determined by mapping reciprocal space around the 200 Bragg peak at 1.8 K.  This peak arises from purely nuclear scattering as the magnetic structure factor is zero at this position.  An estimate for the energy resolution for elastic scattering was determined at a position where only incoherent scattering is expected, corresponding to a rotation from the 200 peak by $15^{\circ}$ about the normal to the scattering plane.  The resolution was then calculated using the Popovici method \cite{Popovici} in the Rescal5 library for Matlab\textsuperscript{\textregistered} \cite{Rescal}. 

\begin{figure}
  \includegraphics[width=3in]{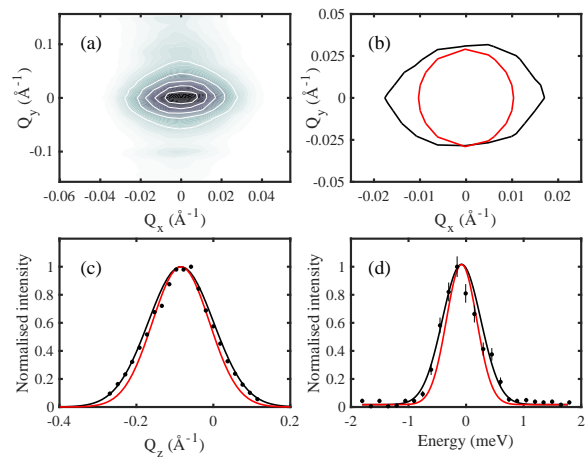}
  \caption{\label{fig:Resolution}Measured and simulated resolution at the 200 position (a) Reciprocal space map of the 200 Bragg peak in the $\left(h0l\right)$ plane. (b) Comparison of the FWHM contour for the measured Bragg peak (black) with the calculated resolution (red). (c) Measured width of the 200 Bragg peak normal to the $\left(h0l\right)$ plane, i.e. along $0k0$.  The fit to the data (black line) and the calculated width (red line) are also shown. (d)  Measured energy width of the incoherent scattering at an equivalent $Q$ to the 200 Bragg peak.  The fit to the data (black line) and the calculated width (red line) are also shown.}	
\end{figure}

The data are shown in figure \ref{fig:Resolution}, along with the estimates from the calculation, with {\bf{Q}} given in reciprocal {\AA}ngstr{\"o}ms.  Figure \ref{fig:Resolution}(a) shows a reciprocal space map of the 200 Bragg peak in the $\left(h0l\right)$ plane, with $Q_x$  and $Q_y$ being parallel and perpendicular to {\bf{Q}} respectively.  The data show that the crystal is not perfect, with the mosaic spread giving rise to a ridge of intensity along $Q_y$.  Figure \ref{fig:Resolution}(b) shows the FWHM contour for the measured peak along with the FWHM for the calculated resolution.  The instrument and measurement parameters for the calculated resolution were all correct and unadjusted for IN8, and the calculation used a mosaic spread of $90^{\prime}$ for the sample.  The comparison is satisfactory in $Q_y$ but is slightly too small along $Q_x$.  The disagreement is not expected to be important at the Brillouin zone boundaries where the magnons have very little dispersion.   Figure \ref{fig:Resolution}(c) shows the width of the 200 peak along $Q_z$, measured by tilting the sample about the $y$-axis.  The peak was fitted with a Gaussian, shown as a black line, which agrees well with the calculated profile, shown as a red line.

The incoherent energy width, shown in figure \ref{fig:Resolution}, is of primary importance.  The measurement here is analogous to a constant {\bf{Q}} scan on a dispersionless mode at zero energy transfer, while the data in figure \ref{fig:Data}(b) come from similar scans of a dispersionless mode at $\sim 11.5$ meV.  The data were fitted with a Gaussian, shown as the black line, while the red line shows the results of the calculation.  The FWHM were $0.79\left(4\right)$ and 0.62 for the fitted and calculated peaks respectively, thus the calculation slightly underestimates the width.  

A similar calculation at $\hbar\omega = 11.5$ meV gives a FWHM of 0.9145 meV, with the widths being very slightly $Q$-dependent in the fourth decimal place.  This may be compared to the fitted FWHM in table \ref{tab:Results}, showing that the values at $\left|\frac{4}{3}0l\right|$ are larger than the calculation by the same magnitude as those at zero energy transfer.  If the difference between fit and calculation is assumed to be systematic, the FWHM of the peaks at the $J$ and $N$ Brillouin zone corners are resolution-limited.  It is worth restating that the splitting observed in the neutron spin-echo spectroscopy measurements was 64 $\mu$eV \cite{Hicks}, which so small as to give a resolution-limited single peak in the three-axis experiment.

\begin{figure}
   \includegraphics[width=3in]{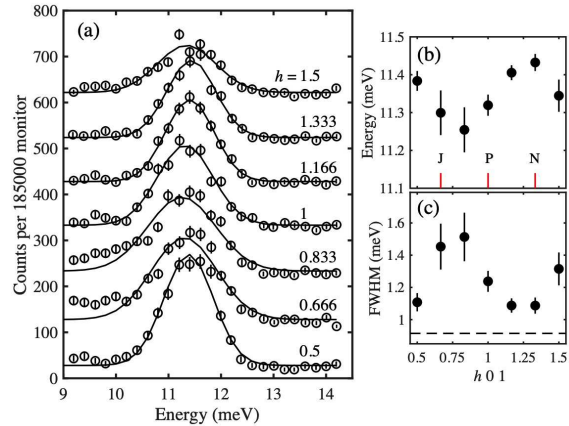}
   \caption{\label{fig:BZB} (a) Measured data along the Brillouin zone boundary along the $h01$ direction, corresponding to the trajectory shown by the red dashed line in fig. \ref{fig:Data}(a).  Data at increasing $h$ are shifted vertically by 100 for clarity.  Fits of Gaussians, including a flat background, to the data are also shown. (b) The fitted centres of the Gaussians as a function of $h$.  The high-symmetry points along the zone boundary are indicated.  (c) The full-width half-maxima of the fitted Gaussians as a function of $h$.  The calculated resolution width for a dispersionless mode at 11.5 meV is shown as the dashed line.}
\end{figure}

Interestingly, a bigger difference is seen at the $P$ position.  The magnons at the six $P$ positions are all expected to have the same energy in the Dzyaloshinskii-Moriya theory and so the widths should be resolution-limited \cite{Cheng}.  The dipole-dipole and bond-specific models predict a splitting of the magnon degeneracy at these positions, but it is much smaller than that predicted at the $J$ points \cite{Pich95, Matsumoto}.  Furthermore, the spin-echo measurements at this position showed a splitting of 39.4 $\mu$eV \cite{Hicks}, much smaller than the IN8 instrument resolution.  The difference may be due to a small amount of suprious scattering on the low-energy side of the peak.  Measurements along the $h01$ direction are shown in figure \ref{fig:BZB}(a).  The data were fitted with Gaussians on a flat background and the fitted centres and FWHM shown in figure \ref{fig:BZB}(b) and (c) respectively.  Some spurious scattering is clearly visible in the lower energies of the data at $h = \frac{2}{3}$ and $\frac{5}{6}$, causing the fitted centres of these peaks to be smaller and the widths to be broader than the neighbours.  Fitting the data with a better estimation of the instrument background, supported by the measurement of spin waves at equivalent points in other Brillouin zones \cite{Wildes98}, show that the actual spin wave energies along this trajectory are practically dispersionless.  The peak at $101$ may also be very slightly impacted, explaining the slightly broader scattering at this point.

Experimental evidence for a magnon Nerst effect in MnPS$_3$ has been published in the form of thermoelectric measurements of bulk crystals with deposited platinum electrodes \cite{Shiomi}.  The evidence was most clearly seen in the temperature-dependence of the thermoelectric coefficient of one of the two electrodes on a sample measured in the absence of a magnetic field.  The coefficient changed sign at $\approx$ 25 K, which would correspond to a Dzyaloshinskii-Moriya parameter of $\approx$ 0.3 meV.  The spin-wave energies described here would suggest that, if present, the Dzyaloshinskii-Moriya parameter must be substantially less than 0.01 meV \cite{Cheng}.  Assuming that the experimental data from the thermoelectric measurements are representative of MnPS$_3$, the observed Nernst effect must depend upon other processes such as magnon-phonon coupling.

In conclusion, neutron three-axis spectroscopy was used to search for non-reciprocal magnons in the Brillouin zone corners of MnPS$_3$.  The data show no convincing evidence for non-reciprocal magnons within the energy resolution of  $\sim 0.9$ meV.   



\begin{acknowledgments}
We thank the Institut Laue-Langevin for the allocation of neutron beam time, Dr. A. Piovano and Dr. A. Ivanov for their assistance with the IN8 instrument, and F. Charpenay and V. Gaignon for technical assistance.  ARW thanks Dr. V. Simonet for a critical reading of the manuscript and Dr. T. Ziman for stimulating discussions.  S.O. was supported by the US Department of Energy, Office of Science, Basic Energy Sciences, Materials Sciences and Engineering Division.  D.X. is supported by AFOSR MURI 2D MAGIC (FA9550-19-1-0390).
\end{acknowledgments}

\bibliography{MPS3}

\begin{thebibliography}{25}%
\makeatletter
\providecommand \@ifxundefined [1]{%
 \@ifx{#1\undefined}
}%
\providecommand \@ifnum [1]{%
 \ifnum #1\expandafter \@firstoftwo
 \else \expandafter \@secondoftwo
 \fi
}%
\providecommand \@ifx [1]{%
 \ifx #1\expandafter \@firstoftwo
 \else \expandafter \@secondoftwo
 \fi
}%
\providecommand \natexlab [1]{#1}%
\providecommand \enquote  [1]{``#1''}%
\providecommand \bibnamefont  [1]{#1}%
\providecommand \bibfnamefont [1]{#1}%
\providecommand \citenamefont [1]{#1}%
\providecommand \href@noop [0]{\@secondoftwo}%
\providecommand \href [0]{\begingroup \@sanitize@url \@href}%
\providecommand \@href[1]{\@@startlink{#1}\@@href}%
\providecommand \@@href[1]{\endgroup#1\@@endlink}%
\providecommand \@sanitize@url [0]{\catcode `\\12\catcode `\$12\catcode
  `\&12\catcode `\#12\catcode `\^12\catcode `\_12\catcode `\%12\relax}%
\providecommand \@@startlink[1]{}%
\providecommand \@@endlink[0]{}%
\providecommand \url  [0]{\begingroup\@sanitize@url \@url }%
\providecommand \@url [1]{\endgroup\@href {#1}{\urlprefix }}%
\providecommand \urlprefix  [0]{URL }%
\providecommand \Eprint [0]{\href }%
\providecommand \doibase [0]{https://doi.org/}%
\providecommand \selectlanguage [0]{\@gobble}%
\providecommand \bibinfo  [0]{\@secondoftwo}%
\providecommand \bibfield  [0]{\@secondoftwo}%
\providecommand \translation [1]{[#1]}%
\providecommand \BibitemOpen [0]{}%
\providecommand \bibitemStop [0]{}%
\providecommand \bibitemNoStop [0]{.\EOS\space}%
\providecommand \EOS [0]{\spacefactor3000\relax}%
\providecommand \BibitemShut  [1]{\csname bibitem#1\endcsname}%
\let\auto@bib@innerbib\@empty
\bibitem [{\citenamefont {Brec}(1986)}]{Brec}%
  \BibitemOpen
  \bibfield  {author} {\bibinfo {author} {\bibfnamefont {R.}~\bibnamefont
  {Brec}},\ }\href@noop {} {\bibfield  {journal} {\bibinfo  {journal} {Solid
  State Ionics}\ }\textbf {\bibinfo {volume} {22}},\ \bibinfo {pages} {3}
  (\bibinfo {year} {1986})}\BibitemShut {NoStop}%
\bibitem [{\citenamefont {Grasso}\ and\ \citenamefont
  {Silipigni}(2002)}]{Grasso}%
  \BibitemOpen
  \bibfield  {author} {\bibinfo {author} {\bibfnamefont {V.}~\bibnamefont
  {Grasso}}\ and\ \bibinfo {author} {\bibfnamefont {L.}~\bibnamefont
  {Silipigni}},\ }\href@noop {} {\bibfield  {journal} {\bibinfo  {journal}
  {Riv. Nuovo Cimento}\ }\textbf {\bibinfo {volume} {25}},\ \bibinfo {pages}
  {1} (\bibinfo {year} {2002})}\BibitemShut {NoStop}%
\bibitem [{\citenamefont {Susner}\ \emph {et~al.}(2017)\citenamefont {Susner},
  \citenamefont {Chyasnavichyus}, \citenamefont {McGuire}, \citenamefont
  {Ganesh},\ and\ \citenamefont {Maksymovych}}]{Susner}%
  \BibitemOpen
  \bibfield  {author} {\bibinfo {author} {\bibfnamefont {M.~A.}\ \bibnamefont
  {Susner}}, \bibinfo {author} {\bibfnamefont {M.}~\bibnamefont
  {Chyasnavichyus}}, \bibinfo {author} {\bibfnamefont {M.~A.}\ \bibnamefont
  {McGuire}}, \bibinfo {author} {\bibfnamefont {P.}~\bibnamefont {Ganesh}},\
  and\ \bibinfo {author} {\bibfnamefont {P.}~\bibnamefont {Maksymovych}},\
  }\href@noop {} {\bibfield  {journal} {\bibinfo  {journal} {Advanced
  Materials}\ }\textbf {\bibinfo {volume} {29}},\ \bibinfo {pages} {1602852}
  (\bibinfo {year} {2017})}\BibitemShut {NoStop}%
\bibitem [{\citenamefont {Wang}\ \emph {et~al.}(2018)\citenamefont {Wang},
  \citenamefont {Shifa}, \citenamefont {Yu}, \citenamefont {He}, \citenamefont
  {Liu}, \citenamefont {Wang}, \citenamefont {Wang}, \citenamefont {Zhan},
  \citenamefont {Lou}, \citenamefont {Xia},\ and\ \citenamefont {He}}]{Wang}%
  \BibitemOpen
  \bibfield  {author} {\bibinfo {author} {\bibfnamefont {F.}~\bibnamefont
  {Wang}}, \bibinfo {author} {\bibfnamefont {T.~A.}\ \bibnamefont {Shifa}},
  \bibinfo {author} {\bibfnamefont {P.}~\bibnamefont {Yu}}, \bibinfo {author}
  {\bibfnamefont {P.}~\bibnamefont {He}}, \bibinfo {author} {\bibfnamefont
  {Y.}~\bibnamefont {Liu}}, \bibinfo {author} {\bibfnamefont {F.}~\bibnamefont
  {Wang}}, \bibinfo {author} {\bibfnamefont {Z.}~\bibnamefont {Wang}}, \bibinfo
  {author} {\bibfnamefont {X.}~\bibnamefont {Zhan}}, \bibinfo {author}
  {\bibfnamefont {X.}~\bibnamefont {Lou}}, \bibinfo {author} {\bibfnamefont
  {F.}~\bibnamefont {Xia}},\ and\ \bibinfo {author} {\bibfnamefont
  {J.}~\bibnamefont {He}},\ }\href@noop {} {\bibfield  {journal} {\bibinfo
  {journal} {Advanced Functional Materials}\ }\textbf {\bibinfo {volume}
  {28}},\ \bibinfo {pages} {1802151} (\bibinfo {year} {2018})}\BibitemShut
  {NoStop}%
\bibitem [{\citenamefont {{Mohamad Latiff}}\ \emph {et~al.}(2019)\citenamefont
  {{Mohamad Latiff}}, \citenamefont {Rosli}, \citenamefont {Mayorga-Martinez},
  \citenamefont {Szokolava}, \citenamefont {Sofer}, \citenamefont {Fisher},\
  and\ \citenamefont {Pumera}}]{Mohamad}%
  \BibitemOpen
  \bibfield  {author} {\bibinfo {author} {\bibfnamefont {N.}~\bibnamefont
  {{Mohamad Latiff}}}, \bibinfo {author} {\bibfnamefont {N.~F.}\ \bibnamefont
  {Rosli}}, \bibinfo {author} {\bibfnamefont {C.~C.}\ \bibnamefont
  {Mayorga-Martinez}}, \bibinfo {author} {\bibfnamefont {K.}~\bibnamefont
  {Szokolava}}, \bibinfo {author} {\bibfnamefont {Z.}~\bibnamefont {Sofer}},
  \bibinfo {author} {\bibfnamefont {A.~C.}\ \bibnamefont {Fisher}},\ and\
  \bibinfo {author} {\bibfnamefont {M.}~\bibnamefont {Pumera}},\ }\href
  {https://doi.org/https://doi.org/10.1016/j.flatc.2019.100134} {\bibfield
  {journal} {\bibinfo  {journal} {FlatChem}\ }\textbf {\bibinfo {volume}
  {18}},\ \bibinfo {pages} {100134} (\bibinfo {year} {2019})}\BibitemShut
  {NoStop}%
\bibitem [{\citenamefont {Park}(2016)}]{Park}%
  \BibitemOpen
  \bibfield  {author} {\bibinfo {author} {\bibfnamefont {J.~G.}\ \bibnamefont
  {Park}},\ }\href@noop {} {\bibfield  {journal} {\bibinfo  {journal} {J.
  Phys.: Condens. Matter}\ }\textbf {\bibinfo {volume} {28}},\ \bibinfo {pages}
  {301001} (\bibinfo {year} {2016})}\BibitemShut {NoStop}%
\bibitem [{\citenamefont {Ouvrard}\ \emph {et~al.}(1985)\citenamefont
  {Ouvrard}, \citenamefont {Brec},\ and\ \citenamefont {Rouxel}}]{Ouvrard85}%
  \BibitemOpen
  \bibfield  {author} {\bibinfo {author} {\bibfnamefont {G.}~\bibnamefont
  {Ouvrard}}, \bibinfo {author} {\bibfnamefont {R.}~\bibnamefont {Brec}},\ and\
  \bibinfo {author} {\bibfnamefont {J.}~\bibnamefont {Rouxel}},\ }\href@noop {}
  {\bibfield  {journal} {\bibinfo  {journal} {Mater.\ Res.\ Bull.}\ }\textbf
  {\bibinfo {volume} {20}},\ \bibinfo {pages} {1181} (\bibinfo {year}
  {1985})}\BibitemShut {NoStop}%
\bibitem [{\citenamefont {Okuda}\ \emph {et~al.}(1986)\citenamefont {Okuda},
  \citenamefont {Kurosawa}, \citenamefont {Saito}, \citenamefont {Honda},
  \citenamefont {Yu},\ and\ \citenamefont {Date}}]{Okuda86}%
  \BibitemOpen
  \bibfield  {author} {\bibinfo {author} {\bibfnamefont {K.}~\bibnamefont
  {Okuda}}, \bibinfo {author} {\bibfnamefont {K.}~\bibnamefont {Kurosawa}},
  \bibinfo {author} {\bibfnamefont {S.}~\bibnamefont {Saito}}, \bibinfo
  {author} {\bibfnamefont {M.}~\bibnamefont {Honda}}, \bibinfo {author}
  {\bibfnamefont {Z.}~\bibnamefont {Yu}},\ and\ \bibinfo {author}
  {\bibfnamefont {M.}~\bibnamefont {Date}},\ }\href@noop {} {\bibfield
  {journal} {\bibinfo  {journal} {J. Phys. Soc. Japan}\ }\textbf {\bibinfo
  {volume} {55}},\ \bibinfo {pages} {4456} (\bibinfo {year}
  {1986})}\BibitemShut {NoStop}%
\bibitem [{\citenamefont {Joy}\ and\ \citenamefont {Vasudevan}(1992)}]{Joy92}%
  \BibitemOpen
  \bibfield  {author} {\bibinfo {author} {\bibfnamefont {P.~A.}\ \bibnamefont
  {Joy}}\ and\ \bibinfo {author} {\bibfnamefont {S.}~\bibnamefont
  {Vasudevan}},\ }\href@noop {} {\bibfield  {journal} {\bibinfo  {journal}
  {Phys. \ Rev. \ B}\ }\textbf {\bibinfo {volume} {46}},\ \bibinfo {pages}
  {5425} (\bibinfo {year} {1992})}\BibitemShut {NoStop}%
\bibitem [{\citenamefont {Kurosawa}\ \emph {et~al.}(1983)\citenamefont
  {Kurosawa}, \citenamefont {Saito},\ and\ \citenamefont
  {Yamaguchi}}]{Kurosawa83}%
  \BibitemOpen
  \bibfield  {author} {\bibinfo {author} {\bibfnamefont {K.}~\bibnamefont
  {Kurosawa}}, \bibinfo {author} {\bibfnamefont {S.}~\bibnamefont {Saito}},\
  and\ \bibinfo {author} {\bibfnamefont {Y.}~\bibnamefont {Yamaguchi}},\
  }\href@noop {} {\bibfield  {journal} {\bibinfo  {journal} {J. Phys. Soc.
  Japan}\ }\textbf {\bibinfo {volume} {52}},\ \bibinfo {pages} {3919} (\bibinfo
  {year} {1983})}\BibitemShut {NoStop}%
\bibitem [{\citenamefont {Ressouche}\ \emph {et~al.}(2010)\citenamefont
  {Ressouche}, \citenamefont {Loire}, \citenamefont {Simonet}, \citenamefont
  {Ballou}, \citenamefont {Stunault},\ and\ \citenamefont
  {Wildes}}]{Ressouche}%
  \BibitemOpen
  \bibfield  {author} {\bibinfo {author} {\bibfnamefont {E.}~\bibnamefont
  {Ressouche}}, \bibinfo {author} {\bibfnamefont {M.}~\bibnamefont {Loire}},
  \bibinfo {author} {\bibfnamefont {V.}~\bibnamefont {Simonet}}, \bibinfo
  {author} {\bibfnamefont {R.}~\bibnamefont {Ballou}}, \bibinfo {author}
  {\bibfnamefont {A.}~\bibnamefont {Stunault}},\ and\ \bibinfo {author}
  {\bibfnamefont {A.}~\bibnamefont {Wildes}},\ }\href@noop {} {\bibfield
  {journal} {\bibinfo  {journal} {Phys. Rev. B}\ }\textbf {\bibinfo {volume}
  {82}},\ \bibinfo {pages} {100408(R)} (\bibinfo {year} {2010})}\BibitemShut
  {NoStop}%
\bibitem [{\citenamefont {Mermin}\ and\ \citenamefont {Wagner}(1966)}]{Mermin}%
  \BibitemOpen
  \bibfield  {author} {\bibinfo {author} {\bibfnamefont {N.~D.}\ \bibnamefont
  {Mermin}}\ and\ \bibinfo {author} {\bibfnamefont {H.}~\bibnamefont
  {Wagner}},\ }\href {https://doi.org/10.1103/PhysRevLett.17.1133} {\bibfield
  {journal} {\bibinfo  {journal} {Phys. Rev. Lett.}\ }\textbf {\bibinfo
  {volume} {17}},\ \bibinfo {pages} {1133} (\bibinfo {year}
  {1966})}\BibitemShut {NoStop}%
\bibitem [{\citenamefont {Pich}\ and\ \citenamefont {Schwabl}(1995)}]{Pich95}%
  \BibitemOpen
  \bibfield  {author} {\bibinfo {author} {\bibfnamefont {C.}~\bibnamefont
  {Pich}}\ and\ \bibinfo {author} {\bibfnamefont {F.}~\bibnamefont {Schwabl}},\
  }\href@noop {} {\bibfield  {journal} {\bibinfo  {journal} {J. Magn. Magn.
  Mater.}\ }\textbf {\bibinfo {volume} {148}},\ \bibinfo {pages} {30} (\bibinfo
  {year} {1995})}\BibitemShut {NoStop}%
\bibitem [{\citenamefont {Cheng}\ \emph {et~al.}(2016)\citenamefont {Cheng},
  \citenamefont {Okamoto},\ and\ \citenamefont {Xiao}}]{Cheng}%
  \BibitemOpen
  \bibfield  {author} {\bibinfo {author} {\bibfnamefont {R.}~\bibnamefont
  {Cheng}}, \bibinfo {author} {\bibfnamefont {S.}~\bibnamefont {Okamoto}},\
  and\ \bibinfo {author} {\bibfnamefont {D.}~\bibnamefont {Xiao}},\ }\href
  {https://doi.org/10.1103/PhysRevLett.117.217202} {\bibfield  {journal}
  {\bibinfo  {journal} {Phys. Rev. Lett.}\ }\textbf {\bibinfo {volume} {117}},\
  \bibinfo {pages} {217202} (\bibinfo {year} {2016})}\BibitemShut {NoStop}%
\bibitem [{\citenamefont {Matsumoto}\ and\ \citenamefont
  {Hayami}(2020)}]{Matsumoto}%
  \BibitemOpen
  \bibfield  {author} {\bibinfo {author} {\bibfnamefont {T.}~\bibnamefont
  {Matsumoto}}\ and\ \bibinfo {author} {\bibfnamefont {S.}~\bibnamefont
  {Hayami}},\ }\href {https://doi.org/10.1103/PhysRevB.101.224419} {\bibfield
  {journal} {\bibinfo  {journal} {Phys. Rev. B}\ }\textbf {\bibinfo {volume}
  {101}},\ \bibinfo {pages} {224419} (\bibinfo {year} {2020})}\BibitemShut
  {NoStop}%
\bibitem [{\citenamefont {Wildes}\ \emph {et~al.}(1998)\citenamefont {Wildes},
  \citenamefont {Roessli}, \citenamefont {Lebech},\ and\ \citenamefont
  {Godfrey}}]{Wildes98}%
  \BibitemOpen
  \bibfield  {author} {\bibinfo {author} {\bibfnamefont {A.~R.}\ \bibnamefont
  {Wildes}}, \bibinfo {author} {\bibfnamefont {B.}~\bibnamefont {Roessli}},
  \bibinfo {author} {\bibfnamefont {B.}~\bibnamefont {Lebech}},\ and\ \bibinfo
  {author} {\bibfnamefont {K.~W.}\ \bibnamefont {Godfrey}},\ }\href@noop {}
  {\bibfield  {journal} {\bibinfo  {journal} {J. Phys: Condens. Matter}\
  }\textbf {\bibinfo {volume} {10}},\ \bibinfo {pages} {6417} (\bibinfo {year}
  {1998})}\BibitemShut {NoStop}%
\bibitem [{\citenamefont {Hicks}\ \emph {et~al.}(2019)\citenamefont {Hicks},
  \citenamefont {Keller},\ and\ \citenamefont {Wildes}}]{Hicks}%
  \BibitemOpen
  \bibfield  {author} {\bibinfo {author} {\bibfnamefont {T.}~\bibnamefont
  {Hicks}}, \bibinfo {author} {\bibfnamefont {T.}~\bibnamefont {Keller}},\ and\
  \bibinfo {author} {\bibfnamefont {A.}~\bibnamefont {Wildes}},\ }\href
  {https://doi.org/https://doi.org/10.1016/j.jmmm.2018.10.136} {\bibfield
  {journal} {\bibinfo  {journal} {Journal of Magnetism and Magnetic Materials}\
  }\textbf {\bibinfo {volume} {474}},\ \bibinfo {pages} {512 } (\bibinfo {year}
  {2019})}\BibitemShut {NoStop}%
\bibitem [{\citenamefont {Goossens}(2010)}]{Goos10}%
  \BibitemOpen
  \bibfield  {author} {\bibinfo {author} {\bibfnamefont {D.~J.}\ \bibnamefont
  {Goossens}},\ }\href@noop {} {\bibfield  {journal} {\bibinfo  {journal} {Eur.
  Phys. J. B}\ }\textbf {\bibinfo {volume} {78}},\ \bibinfo {pages} {305}
  (\bibinfo {year} {2010})}\BibitemShut {NoStop}%
\bibitem [{\citenamefont {Wildes}\ \emph {et~al.}(2006)\citenamefont {Wildes},
  \citenamefont {R{\o}nnow}, \citenamefont {Roessli}, \citenamefont {Harris},\
  and\ \citenamefont {Godfrey}}]{Wildes06}%
  \BibitemOpen
  \bibfield  {author} {\bibinfo {author} {\bibfnamefont {A.~R.}\ \bibnamefont
  {Wildes}}, \bibinfo {author} {\bibfnamefont {H.~M.}\ \bibnamefont
  {R{\o}nnow}}, \bibinfo {author} {\bibfnamefont {B.}~\bibnamefont {Roessli}},
  \bibinfo {author} {\bibfnamefont {M.~J.}\ \bibnamefont {Harris}},\ and\
  \bibinfo {author} {\bibfnamefont {K.~W.}\ \bibnamefont {Godfrey}},\
  }\href@noop {} {\bibfield  {journal} {\bibinfo  {journal} {Phys. Rev. B}\
  }\textbf {\bibinfo {volume} {74}},\ \bibinfo {pages} {094422} (\bibinfo
  {year} {2006})}\BibitemShut {NoStop}%
\bibitem [{\citenamefont {Wildes}\ \emph {et~al.}(2019)\citenamefont {Wildes},
  \citenamefont {Anand},\ and\ \citenamefont {Xiao}}]{Wildes_IN8_Jun19}%
  \BibitemOpen
  \bibfield  {author} {\bibinfo {author} {\bibfnamefont {A.}~\bibnamefont
  {Wildes}}, \bibinfo {author} {\bibfnamefont {K.}~\bibnamefont {Anand}},\ and\
  \bibinfo {author} {\bibfnamefont {D.}~\bibnamefont {Xiao}},\ }\bibfield
  {title} {\bibinfo {title} {The nature of the anisotropy in {M}n{PS}$_3$}\
  }\href {https://doi.org/doi.ill.fr/10.5291/ILL-DATA.4-01-1607}
  {doi.ill.fr/10.5291/ILL-DATA.4-01-1607} (\bibinfo {year} {2019})\BibitemShut
  {NoStop}%
\bibitem [{\citenamefont {Murayama}\ \emph {et~al.}(2016)\citenamefont
  {Murayama}, \citenamefont {Okabe}, \citenamefont {Urushihara}, \citenamefont
  {Asaka}, \citenamefont {Fukuda}, \citenamefont {Isobe}, \citenamefont
  {Yamamoto},\ and\ \citenamefont {Matsushita}}]{Murayama}%
  \BibitemOpen
  \bibfield  {author} {\bibinfo {author} {\bibfnamefont {C.}~\bibnamefont
  {Murayama}}, \bibinfo {author} {\bibfnamefont {M.}~\bibnamefont {Okabe}},
  \bibinfo {author} {\bibfnamefont {D.}~\bibnamefont {Urushihara}}, \bibinfo
  {author} {\bibfnamefont {T.}~\bibnamefont {Asaka}}, \bibinfo {author}
  {\bibfnamefont {K.}~\bibnamefont {Fukuda}}, \bibinfo {author} {\bibfnamefont
  {M.}~\bibnamefont {Isobe}}, \bibinfo {author} {\bibfnamefont
  {K.}~\bibnamefont {Yamamoto}},\ and\ \bibinfo {author} {\bibfnamefont
  {Y.}~\bibnamefont {Matsushita}},\ }\href@noop {} {\bibfield  {journal}
  {\bibinfo  {journal} {J. Appl. Phys.}\ }\textbf {\bibinfo {volume} {120}},\
  \bibinfo {pages} {142114} (\bibinfo {year} {2016})}\BibitemShut {NoStop}%
\bibitem [{\citenamefont {Ouvrard}\ and\ \citenamefont
  {Brec}(1990)}]{Ouvrard90}%
  \BibitemOpen
  \bibfield  {author} {\bibinfo {author} {\bibfnamefont {G.}~\bibnamefont
  {Ouvrard}}\ and\ \bibinfo {author} {\bibfnamefont {R.}~\bibnamefont {Brec}},\
  }\href@noop {} {\bibfield  {journal} {\bibinfo  {journal} {Eur. J. Solid
  State Inorg. Chem.}\ }\textbf {\bibinfo {volume} {27}},\ \bibinfo {pages}
  {477} (\bibinfo {year} {1990})}\BibitemShut {NoStop}%
\bibitem [{\citenamefont {Popovici}(1975)}]{Popovici}%
  \BibitemOpen
  \bibfield  {author} {\bibinfo {author} {\bibfnamefont {M.}~\bibnamefont
  {Popovici}},\ }\href {https://doi.org/10.1107/S0567739475001088} {\bibfield
  {journal} {\bibinfo  {journal} {Acta Crystallographica Section A}\ }\textbf
  {\bibinfo {volume} {31}},\ \bibinfo {pages} {507} (\bibinfo {year}
  {1975})}\BibitemShut {NoStop}%
\bibitem [{Res()}]{Rescal}%
  \BibitemOpen
  \href@noop {} {}\bibinfo {howpublished}
  {\url{http://ifit.mccode.org/Applications/ResLibCal/doc/ResLibCal.html}},\
  \bibinfo {note} {accessed: 2020-07-02}\BibitemShut {NoStop}%
\bibitem [{\citenamefont {Shiomi}\ \emph {et~al.}(2017)\citenamefont {Shiomi},
  \citenamefont {Takashima},\ and\ \citenamefont {Saitoh}}]{Shiomi}%
  \BibitemOpen
  \bibfield  {author} {\bibinfo {author} {\bibfnamefont {Y.}~\bibnamefont
  {Shiomi}}, \bibinfo {author} {\bibfnamefont {R.}~\bibnamefont {Takashima}},\
  and\ \bibinfo {author} {\bibfnamefont {E.}~\bibnamefont {Saitoh}},\ }\href
  {https://doi.org/10.1103/PhysRevB.96.134425} {\bibfield  {journal} {\bibinfo
  {journal} {Phys. Rev. B}\ }\textbf {\bibinfo {volume} {96}},\ \bibinfo
  {pages} {134425} (\bibinfo {year} {2017})}\BibitemShut {NoStop}%
\end{thebibliography}%

\end{document}